\documentclass[onecolumn,showpacs,amsmath,amssymb,prd,nofootinbib]
{revtex4}
\def\beq{\begin{equation}}
\def\eeqno#1{\label{#1}\end{equation}}
\def\sss{\scriptscriptstyle}
\def\^#1{^{\sss #1}}
\def\_#1{_{\sss #1}}
\def\div{\vec\nabla\cdot}
\def\grad{\vec\nabla}
\def\Grad{{\bf D}}

\def\rar{\rightarrow}
\def\s{\sigma}
\def\z{\zeta}
\def\eps{\epsilon}
\def\a{\alpha}
\def\b{\beta}
\def\l{\lambda}
\def\m{\mu}
\def\n{\nu}
\def\r{\rho}
\def\c{\gamma}

\def\d{\delta}
\def\f{\phi}

\def\L{\mathcal{L}}
\def\J{\mathcal{J}}
\def\U{\mathcal{U}}
\def\S{\mathcal{S}}
\def\I{\mathcal{I}}

\def\H{\mathcal{H}}
\def\bU{\bar{\mathcal{U}}}
\def\bS{\bar{\mathcal{S}}}

\def\emn{\eta\_{\m\n}}

\def\Gmn{g\^{\mu \nu}}

\def\fmn{F_{\m\n}}
\def\Fmn{F^{\mu \nu}}
\def\tFmn{\tilde F^{\mu \nu}}
\def\qmn{Q_{\m\n}}
\def\Qmn{Q^{\mu \nu}}
\def\hmn{H_{\m\n}}
\def\Hmn{H^{\mu \nu}}

\def\tFmn{\tilde F^{\mu \nu}}

\def\tQmn{\tilde Q^{\mu \nu}}

\def\Pmn{P^{\mu \nu}}
\def\oot{\frac{1}{2}}
\def\ddag{d^\dag}
\def\qf{q\_F}
\def\pf{p\_F}
\def\qq{q\_Q}
\def\pq{p\_Q}
\def\gf{\grad\f}
\def\Gf{\Grad\f}
\def\gfs{(\gf)^2}
\def\gps{\grad\psi}

\def\Gps{\Grad\psi}

\def\az{a\_0}
\def\vr{{\bf r}}
\def\vx{{\bf x}}
\def\vy{{\bf y}}
\def\vv{{\bf v}}
\def\ve{{\bf e}}

\def\vA{{\bf A}}
\def\vB{{\bf B}}
\def\vg{{\bf g}}
\def\vf{{\bf f}}
\def\vF{{\bf F}}
\def\vP{{\bf P}}
\def\vE{{\bf E}}

\def\vJ{{\bf J}}
\def\vD{{\bf D}}
\def\vH{{\bf H}}
\def\dvs{d{\vec \sigma}}

\def\tU{\tilde\U}
\def\tS{\tilde\S}
\begin{document}

\title{Practically linear analogs of the Born-Infeld and other nonlinear theories}
\author{Mordehai Milgrom} \affiliation{DPPA, Weizmann Institute of
Science, Rehovot 76100, Israel}

\begin{abstract}
I discuss theories that describe fully nonlinear
physics, while being practically linear (PL), in that they require solving only linear differential equations. These theories may be interesting in themselves as manageable nonlinear theories. But, they can also be chosen to emulate genuinely nonlinear theories of special interest, for which they can serve as approximations. The idea can be applied to a large class of nonlinear theories, exemplified here with a PL analogs of scalar theories, and of Born-Infeld (BI) electrodynamics. The general class of such PL
theories of electromagnetism are governed by a Lagrangian $\L=-(1/2)\fmn\Qmn+ \tS(\qmn)$, where $\fmn=A_{\n,\m}-A_{\m,\n}$, and $A_{\m}$ couples to currents in the standard way, while $\qmn=B_{\n,\m}-B_{\m,\n}$ is an auxiliary field that does not couple directly to currents.
By picking a special form of $\tS(\qmn)$, we can make such a theory similar in some regards to a given fully nonlinear theory, governed by the Lagrangian $-\tU(\fmn)$. For example, by ``similar'' we may imply that the theories are equivalent to second order in the expansion for weak fields, and that they are also equivalent for static configurations with one-dimensional symmetry (e.g., near point charges). A particularly felicitous choice, which implies the above similarities, is to take $\tS$ as the Legendre transform of $\tU$ in the variables $\fmn$. For the BI theory, this Legendre transform has the same form as the BI Lagrangian itself:  $\tS(\qmn,E_0^2)=\tU(\qmn,-E_0^2)$ ($E_0$ is the limiting field of the BI theory).
Various matter-of-principle questions remain to be answered regarding such theories.
As a specific example, I discuss BI electrostatics in more detail. As an aside, for BI, I derive an exact expression for the short-distance force between two arbitrary point charges of the same sign, in any dimension.

\end{abstract}
\maketitle

\section{Introduction}
Nonlinear systems are rife in physics.
My focus here is on theories for which the Lagrangian density is a nonquadratic function of the derivatives of some of the degrees of freedom (DoF).
\par
Some examples of such scalar systems in Euclidean space are:
(i) Electrostatics in nonlinear dielectrics, magnetostatics in the presence of superconductors, or nonlinear transport systems (e.g., nonlinear diffusion).
(ii) Inviscid, irrotational, compressible, stationary flows.
(iii) The problem of volume extremization.
(iv) Alternative theories of gravity replacing the Poisson equation in the description of nonrelativistic gravity by a nonlinear version \cite{bm84}.
\par
Other important examples of nonlinear physics involving vector fields are the different versions of nonlinear electrodynamics, such as that governed by the Heisenberg-Euler Lagrangian (see, e.g., \cite{dunne04}), and the Born-Infeld (BI) theory with its more recent generalizations.
Born-Infeld theories have been repeatedly coming back into the limelights since their advent, almost eighty years ago, because of their unique properties (see, e.g., \cite{gibbons96}\cite{deser99}), and, in particular, they have attracted much attention in recent years, because they appear as effective theories in the context of string theory. (e.g., \cite{tseytlin99}).
\par
Such theories are notoriously unwieldy due to their nonlinearity.
Here I discuss a type of theories that describe nonlinear physics, while being practically linear (PL), requiring solving only linear differential equations, the nonlinearity entering only algebraically.
\par
Even if such PL theories are not forced on us by nature, they may be useful as wieldy NL theories that embody many of the attributes of genuinely nonlinear theories. Furthermore, for a given nonlinear theory of the type focused on here, we can find a kindred among the PL theories that mimics it in some regards, and could thus serve as a useful approximation.
\par
The theories I discuss here came to light as a result of attempting to find approximations for the NL Poisson, modified-gravity theory alluded to in (iv) above. In \cite{milgrom10} I described such a PL theory, called QUMOND; I showed that it can be considered  a full-fledged theory on its own right (there even is a covariant relativistic MOND theory, for which QUMOND is the nonrelativistic limit \cite{milgrom09}), and I showed that it may be so chosen as to approximate the nonlinear Poisson theory in various circumstances. I also discussed some of the differences between the two theories.  This PL version of MOND has since been put to good use for predicting and calculating
MOND effects in the solar system \cite{milgrom09a}\cite{galianni11}, for calculating MOND fields of galaxies (e.g. \cite{angus12}), and  structure formation in MOND \cite{llinares11}. Here, I essentially extend the concept to more general NL problems.
\par
I first demonstrate the idea with theories for scalar fields, in section \ref{scalar}. Section \ref{bi} deals with nonlinear electrodynamics, and PL analogs of Born-Infeld electrodynamics.
In section \ref{rk}, I describe a simple application to special-relativistic particle kinematics.
Section \ref{electrostatics} discusses BI electrostatics in more detail, as an example of an application.
In section \ref{discussion}, I list some of the many aspects that remain to be checked and considered.

\section{\label{scalar}Scalar theories}
Consider a NL theory involving one real, scalar field, $\f$, that is governed by an action of the form
\beq I=-\int dV~ \tU(\f_{,\m}) +I_q, \eeqno{jutces}
where $dV$ is the appropriate volume element. The action $I_q=-\int q\f~dV$ includes the interaction of $\f$ with other DoF, and is assumed to depend linearly on $\f$.
The field equation for $\f$ is then
 \beq  \left(\frac{\partial \tU}{\partial \f_{,\m}}\right)_{,\m}=q,\eeqno{cureda}
 where the charge distribution, $q$, depends on the configuration of other DoF, but is independent of $\f$.
If the theory is rotationally-invariant (or Lorentz-invariant, or diffeomorphism invariant, depending on the background space) the action is of the form
\beq I=-\int dV~ \U(\f_{,\m}\Gmn\f_{,\n}) +I_q.  \eeqno{muver}
Here, $\Gmn$ is the (fixed) metric of the background space (assumed, for simplicity, to be flat Euclidean or Minkowski in what follows).
 The field equation for $\f$ is then
   \beq 2(\U'\Gmn\f_{,\m})_{,\n} =q(x^\m). \eeqno{kipev}
\par
For the nonlinear systems mentioned in the introduction we have:
(i) In nonlinear dielectrics, nonlinear transport systems, etc., $\U'$ is the response coefficient, which is a function of the field strength, and $q$ represents the density of sources.
(ii) In ideal, irrotational, compressible, stationary flow problems, $\U'$ is the fluid density, which can be expressed as a function of the fluid velocity $\vv=\gf$ through the Bernoulli equation;  $q$ is the source density. For example, in a fluid with an equation of state of the form
 $p=a \varrho^{\c}$ ($a>0,~\c\ge 1$), we have
 $\U'(z)\propto
 [1-(z/z_0)^2]^{1/(\c-1)}$, with $z_0^2\equiv 2c_0^2/(\c-1)$, and
$c_0$ is the speed of sound at $z=0$.
(iii) In the problem of volume extremization of an $(N-1)$-dimensional manifold $x_1=\f(x_2,...,x\_N)$, embedded in an $N$-dimensional Euclidean space, we have $\U(z)\propto(1+z^2)^{1/2}$, and $q(x_2,...,x\_N)$ may be understood as the density of external forces in the $x_1$ direction (as for a loaded soap film in a constant gravitational field).
The above theories, as many others, tend to a linear theory in the limit of weak (gradient) fields: in these cases, $\U'(0)$ is a finite constant.
(iv) in MOND gravity, which replaces the Poisson equation for the gravitational potential by a nonlinear version \cite{bm84}, we have $\U(z)\propto\az^2F(z/\az^2)$ ($\az$ is the acceleration constant of the theory).
This theory is unique among the rest presented above in that it tends to the linear Poisson theory in the strong-field limit: $\U'(z\rar\infty)\rar const.$, while in the weak-field limit $\U'(z\ll 1)\propto z^{1/2}$, in order to reproduce galaxy dynamics without ``dark matter''.
\subsection{The practically linear theory}
Introduce the PL analog theory for a single scalar as follows: Start with the action
  \beq I=\int dV~[-\f_{,\m}e^\m +\tS(\ve)] +I_q,  \eeqno{mudfta}
where $\ve$ is an auxiliary, vector DoF.
As before, $I_q$ couples to $\f$ (linearly), but not to $\ve$.
We then get the field equations
 \beq e^\m_{,\m}=q,~~~~~\f_{,\m}=\frac{\partial \tS}{\partial e^\m}.  \eeqno{kizbuta}
If the Hessian of $\tS$ is regular (needed for an acceptable theory, and assumed all along) the second set of equations may be inverted to give $\ve(\Grad\f)$ ($\Grad$ is the gradient). Furthermore, it can be shown that this inversion involves a single function $\tU(\Gf)$
such that\footnote{Define $q_\a=\partial\tS/\partial e^\a$, then, taking the partial derivative of this relation with respect to $q_\b$ gives $\d^\b_\a=(\partial^2\tS/\partial e^\a\partial e^\c)(\partial e^\c/\partial q_\b)$. Multiplying by the inverse of the Hessian of $\tS$, which is a symmetric metric, we see that $\partial e^\a/\partial q_\b$ is symmetric under the interchange of $\b$ and $\a$. This means that there exists a function $\tU(\vec q)$ such that $e^\a=\partial \tU/\partial q_\a$.}
 \beq e^{\m}=\frac{\partial \tU}{\partial \f_{,\m}}. \eeqno{gukida}
$\tU$ is such that its Hessian and that of $\tS$ are mutual inverses. Substituting in the first of equations (\ref{kizbuta}), we see that $\f$ satisfies the field equation(\ref{cureda}).
\par
If we substitute in the Lagrangian in eq.(\ref{mudfta}) $\Gf$ from the second of eq(\ref{kizbuta}), and express the resulting Lagrangian in terms of $\Gf$, we get, by definition, minus the Legendre transform of $\tS(e^{\m})$. But this can be seen to equal $-\tU(\Gf)$ up to a constant (because its derivative with respect to $\f_{,\m}$ is $e^\m)$.
In other words, the theory (\ref{mudfta}) with $\ve$ an independent DoF, is equivalent to the NL scalar theory (\ref{jutces}) with $\tU$ the Legendre transform of $\tS$.
\par
If, however, we do not permit $\ve$ to be a general vector field, but constrain it, a priori, to be a gradient, $\ve=\Gps$, with $\psi$ as independent DoF, we get a theory of the PL type.
It is governed by the action
 \beq I=\int dV [-\Gf\cdot\Gps +\tS(\Gps)] +I_q,  \eeqno{mudkla}
(dot product with the appropriate metric, which also raises and lowers indices) and it's field equations are
 \beq \Box\psi=q,~~~~~~~~~\Box\f=
 \left(\frac{\partial\tS}{\partial\psi_{,\m}}\right)_{,\m}\equiv q\_{\f}. \eeqno{kuoshla}
In the rotationally invariant case we write the action and the field equations as:
 \beq I=\int dV~\{-\Gf\cdot\Gps +\S[(\Gps)^2]\} +I_q,  \eeqno{mudder}
 \beq \Box\psi=q,~~~~~~~~\Box\f=2(\S'\psi_{,}^{\m})_{,\m}\equiv q\_{\f}.  \eeqno{megna}
\par
To solve these equations we need first to solve the linear equation for $\psi$, with the source distribution $q(x^\m)$, with the appropriate boundary conditions (BC). Then, substitute the solution in the expression for $q\_{\f}(x^\m)$, which becomes a nonlinear, and nonlocal, functional of $q(x^\m)$. Then solve the linear equation for $\f$, with $q\_{\f}$ as source. This involves solving only the linear (Poisson) equation twice, with an algebraic step in between. So the practical advantages of such theories are obvious.
\par
The pair of equations (\ref{kizbuta}) also look like two linear
equations. But this is an illusion: one cannot, of course, solve the first for $e_\m$, and then substitute in the second and solve for $\f$ (for example, the first equation determines $e_\m$ only up to a divergenceless field). It is only when we constrain $\ve$ to be a gradient that the first equation does determine it (given the appropriate BC).
\par
If we want to approximate a given NL theory governed by some $\tU(\Gf)$, with a PL theory, then it would be a good choice to take $\tS$ of the PL theory to be the Legendre transform of $\tU$. This choice automatically guarantees certain similarities between the two theories.
In the first place it guarantees coincidence of the solutions in cases of 1-D symmetry\footnote{For example spherically- or cylindrically-symmetric configurations, if we work in Euclidean space, or static, spherically-symmetric configurations in Minkowski space-time.}, such as near point or line charges: in such cases all vector fields are gradients, and so the gradient constraint on $\ve$, which was imposed to get the PL theory, is automatically satisfied; so the solution of the PL theory
is automatically the solution of the NL theory (for the same BC).
\par
Secondly, as I show below in section \ref{multi}, such a choice of $\tS$ guarantees that for weak fields, the solutions of the two theories coincide to the next order above the lowest, linear case.
\par
Defining the Lagrangian of the PL theory is not enough to fix the theory.
There remains the issue of what BC we dictate for $\psi$. While $\f$ is the ``physical'' field that is felt by charges directly, so we usually know what BC we want for it, $\psi$ is auxiliary, and, in principle, we may have more freedom in choosing its BC. The procedure of picking BC for $\psi$ is part and parcel of the theory:
Different choices of the BC of $\psi$ lead to different solutions for $\f$ even for the same BC. Sometimes, however, the BC for $\psi$ as well, are dictated by the problem. For example, in the problem of BI electrostatics, to be discussed in more detail below, or in the case of the NL, nonrelativistic MOND, Gauss's theorem, with the requirement that the solution becomes symmetric at infinity
(spherically, cylindrically, etc.), applied to the first of eqs.(\ref{megna}) implies that $\psi\rar 0$ at spatial infinity (e.g., $\gps\propto \vr/r^3$ in the 3-D spherical case).
More generally this may be a thorny issue that may incapacitate the method in some cases.
\par
The fully NL theory has a linear, weak-field limit, if $\partial\tU/\partial\f_{,\m}(0)=0$, and $\partial^2\tU/\partial\f_{,\m}\partial\f_{,\n}(0)=\d^{\m\n}$, and similarly for the PL theory (for $|\f_{,\m}|,~|\psi_{,\m}|\ll 1$).
In the weak-field limit, the action of the PL theory becomes
 \beq I\approx \int dV~[-\oot(\Gf)^2 +\oot(\Grad\chi)^2]+I_q,  \eeqno{mufet}
where $\chi=\f-\psi$.
We see that if we work in Minkowski space-time, in which case our metric convention is $\emn=(-1,1,1,1)$, the kinetic term for $\f$ has the ``correct'' sign corresponding to positive kinetic energy, but $\chi$  has the ghost-like sign of the free Lagrangian, but it decouples from all other DoF in this limit. So, the standard, linear theory is gotten.
\par
The ghost-like nature of $\chi$ might bode trouble for the full theory, when time dependent problems are considered. In stationary problems, such as all the examples above, this is not an issue. Even for a fully dynamical situations it is not clear that this aspect is deleterious, since $\chi$ is not quite an independent degree of freedom that may be manipulated in itself. While $\chi$ waves may be carrying negative energy to infinity (where the linear approximation is good), it is not clear that there are charge configurations that emit net negative energies in toto, and thus become unacceptably unstable.
This is, however, an important concern that remains to be addressed.
\par
As was discussed in \cite{milgrom10}, the relation demonstrated here between the pair of theories is analogous to that between the standard and the Palatini formulations of gravitational, metric theories. The Palatini-like approach, whereby $\ve$ is independent (i.e., not assumed to be a gradient) gives the NL theory. The ``standard'' approach, with $\ve$ a gradient, yields the PL theory. For the linear case, both routes of variation give the same field equation as the second of equations (\ref{kizbuta}) becomes $\ve=\Gf$.

\subsection{Similarity in more detail}
\subsubsection{Equivalence for one dimensional configuration}
Consider in some more detail, the case of the rotationally-invariant theory. So $\tS=\S(e^2)$, and $\tU=\U(\Gf^2)$.
Consider a 1-D-symmetric configuration, such as a spherical symmetry in flat space. If the problem at hand is posed in a Minkowskian space, assume that the configuration is also static so the problem can be posed in the Euclidean space.
Let the symmetry surfaces be designated by the coordinate $r$, so that $q=q(r)$. By applying Gauss's theorem to both theories for a volume within a constant $r$, we see that the gradient of the potential $\f$ is a function of only $Q(r)$, the charge enclosed within $r$. The form of this function depends on $\U$ and $\S$, respectively, for the two theories. So, given $\U$ we can choose $\S$ such that $\f$ will depend on $Q(r)$ in the same way in both theories. It is easily seen that the condition for this is as follows: If we define the variable $y$ such that\footnote{$z$ stands for $|\Gf|^2$, and $y$ stands for $|\Gps|^2$.}
 \beq y^{1/2}=2\U'(z)z^{1/2}, \eeqno{nutesa}
 then $\S$ has to satisfy
 \beq z^{1/2}=2\S'(y)y^{1/2}. \eeqno{nutenu}
Either equation has to define $y$ and $z$ as monotonic functions of one another.
 \par
It is easy to see that requirement (\ref{nutesa})(\ref{nutenu}) is tantamount to  $\tS$ and $\tU$ being mutual Legendre transforms (in the components of $\ve$ and $\Gf$ as variables, not in their squares), with the Hessians of $\tU$ and $\tS$ mutual inverses\footnote{The product of the Hessians is $4\S'\U'\d_{\m\n}+e_\m e_\n(32\S'^3\U''+8\S''\U'+64\S''\U''\S'^2e^2)$, and is seen to give $\d_{\m\n}$, from eqs.(\ref{nutesa})(\ref{nutenu}).}.
\par
So in this case equivalence for 1-D configurations uniquely determines $\tS$  to be the Legendre transform of $\tU$ by the similarity requirements.
This is not so for more general cases, such as multi-scalar theories, or the BI theory (see below).
\subsubsection{Equivalence to second order in weak fields}
Suppose the nonlinear theory has a linear weak-field limit
($|\Gf|\ll 1$); so we have $\U'(0)=1/2$. The PL theory has this limit if $\S'(0)=1/2$.
In this limit $\Box\bar\f=\Box\bar\psi=-q$, $\Box\bar\chi=0$ (where a bar designates the solution of the linear theory for the same charge distribution and BC). Writing  $\f=\bar\f+\eta$, we have to lowest order in $\eta$
\beq \Box\eta=-2\U''(0)[(\Grad\bar\f)^2\bar\f_,^{\m}]_{,\m},  \eeqno{kipoda}
in the NL theory, and
in the PL theory we have ($\bar\psi=\bar\f$)
\beq \Box\eta=2\S''(0)[(\Grad\bar\f)^2\bar\f_,^{\m}]_{,\m}.  \eeqno{kipura}
The two are the same if  $\S''(0)=-\U''(0)$. If $\U''(0)=0$, there is a similar condition on the first derivative that does not vanish at zero (see below in \ref{multi}).
\par
Condition (\ref{nutenu}) guarantees this: it implies that $4\U'(z)\S'(y)=1$ for $y$ and $z$ related by eq.(\ref{nutesa}).
Taking the $y$ derivative of this at $y=z=0$ we get $\S''(0)=-\U''(0)$. So the equivalence of the theories for 1-D configurations implies equivalence to second order for all configurations.
\subsection{Weak perturbations in the two theories}
Suppose the solutions of the field equations of the two theories $\bar\f$, $\bar\psi$ are known for a charge distribution $\bar q$. Expanding about this solution to first order in a small perturbation $\eta=\f-\bar\f$, $\l=\psi-\bar\psi$, caused by a small change in the density $\eps=q-\bar q$, we have for the NL theory
 \beq [\H^{\m\n}_u(\Grad\bar\f)\eta_{,\m}]_{,\n}=\eps, \eeqno{mmagopaq}
 and for the kindred PL theory
 \beq \Box\l=\eps,~~~~~~
 \Box\eta=[\H^{\m\n}_s(\Grad\bar\psi)\l_{,\m}]_{,\n}\eeqno{jioter}
where $\H_u$ and $\H_s$ are the Hessians of $\tU$ and $\tS$ respectively. These two sets of field equations are, generally, not strongly related. However, as we saw, if the unperturbed problem is of 1-D symmetry,  $\H_u$ and $\H_s$ are mutual inverses everywhere; so the two perturbation problems become related (but not the same): The NL theory gives an analog of scalar, linear electrodynamics in a position-dependent, anisotropic dielectric, which we can write as
\beq [A^{\m\n}(x)\eta_{,\m}]_{,\n}=\Box\l,\eeqno{nashn}
which is still difficult to solve, generally. The PL theory gives the Poisson equation
\beq\Box\eta=[{(A^{-1})}^{\m\n}(x)\l_{,\m}]_{,\n}, \eeqno{gumpert}
 where $\Box\l=\eps$.
 \par
In some instances we deal with a charge system $\eps$, of small extent, embedded in a meta-system, whose effect on the subsystem may be approximated by a constant external field. This external field is then not part of the dynamics, but is dictated as BC: We seek to solve the NL or PL problem for $\eps$, where in the former we dictate the BC of constant
$\Gf=\vg_0$ at infinity, while in the latter we have to dictate both $\Gf=\vg_0$ and $\Grad\psi=\vf_0$ at infinity. In this latter case $\vg_0$ and $\vf_0$ are not a priori related without specifying what the meta-system is like, and where in it the subsystem is. For example, if the meta-system has 1-D symmetry, $\vg_0$ and $\vf_0$ are parallel and their magnitudes are related.
If, in addition, the subsystem can be treated as a small perturbation, we have eqs.(\ref{nashn})(\ref{gumpert}), with $A^{\m\n}$ now a constant matrix. For example, when $\tU(\Gf)=\U[(\Gf)^2]$ (and similarly for $\tS$), we have
\beq A=\H_u=2\U'_0(I+2\hat\U'_0 \ve_0\otimes\ve_0), \eeqno{netraf}
where $I$ is the unit matrix, $\U'_0$ and $\hat\U'_0$ are the values of $\U'$ and its logarithmic derivative calculated at $\vg_0$, and $\ve_0$ is a unit vector in the direction of $\vg_0$.
Taking, say, the 1 axis in the direction of $\ve_0$, and defining $\hat\l=\l/2\U'_0$ we can write eq.(\ref{nashn}) as
 \beq \Box\eta+2\hat\U'_0\eta_{,1,1}
 =\Box\hat\l,\eeqno{mamauta}
and eq.(\ref{gumpert}) as
\beq \Box\eta =[(1+2\hat\U'_0)^{-1}-1]\hat\l_{,1,1}+\Box\hat\l,
\eeqno{kikilat}
with $\Box\hat\l=\eps/2\U'_0$. In the coordinates $\bar x^1=(1+2\hat\U'_0)^{-1/2}x^1$, $\bar x^i=x^i$ for $i>1$, eq(\ref{mamauta}) takes the same form as eq.(\ref{kikilat}), and both theories then involve solving a linear Poisson equation.
\subsection{\label{multi}Multi-scalar theories}
Consider a NL theory of many scalar fields, with the Lagrangian:
 \beq \L=-\tU(\Gf^1,...,\Gf\^N)-\sum_{a=1}\^N\f^aq^a,  \eeqno{cytreq}
leading to the field equations
\beq\left(\frac{\partial\tU}{\partial\f_{,\m}^a}\right)_{,\m}=q^a.
\eeqno{nuged}
As in the single-scalar case, the Lagrangian
\beq \L=\sum_a-\Gf^a\cdot\ve^a+\tS(\ve^1,...,\ve\^N)-\sum_{a=1}\^N\f^aq^a,  \eeqno{cytbla}
with $\tS$ the Legendre transform of $\tU$ (with respect to all variables), gives the field equations (\ref{nuged}) for $\f^a$, and is an equivalent theory. If, however, we constrain $\ve^a$ to be gradient fields, $\ve^a=\Gps^a$, with $\psi^a$ the fundamental DoF, we get a different theory: the PL theory, whose field equations are:
 \beq \Box\psi^a=q^a,~~~~~~~~~\Box\f^a=
 \left(\frac{\partial\tS}{\partial\psi_{,\m}^a}\right)_{,\m}\equiv q^a_\f. \eeqno{kuopit}
For 1-D configurations, all the $\ve^a$ are automatically (parallel) gradients; so the constraint leading to  eqs.(\ref{kuopit}) is anyhow satisfied even in the NL theory; so the solutions of the two theories coincide.
In this context, there is a difference between the single- and multi-scalar theories: In the single-scalar, rotationally-invariant theory (\ref{muver}), $\tU$ and $\tS$ are functions of only one variable, as they depend on $\Gf$, or $\Gps$, through $(\Gf)^2$, or $(\Gps)^2$, respectively. Then, the requirement of coincidence for 1-D configurations is enough to pinpoint $\tS$, uniquely, as the Legendre transform of $\tU$. For a multi-scalar theory, this is not the case, even for rotationally-invariant theories: Now, $\tU$ and $\tS$ depend on their vector variables through the invariants, $\Gf^a\cdot\Gf^b$, or $\Gps^a\cdot\Gps^b$. But for 1-D configurations, only a subset of the variable values is probed, since each of the $\Gf$ and $\Gps$ is determined by only one component, and $\tS$ and $\tU$ become functions of only these $N$ single components. So, clearly, only the dependence of $\tS$ on a subset of its variables enters, and is constrained, by the requirement of 1-D equivalence. This, as we shall see, is the case for NL electromagnetism as well.
The Legendre-transform choice is thus not unique. But it might have additional, yet unappreciated advantages.
\par
Another attraction of taking $\tS$ to be the Legendre transform of $\tU$, is that it gives a PL theory that coincides to next to leading order in weak-fields, with the NL $\tU$ theory:
Suppose the NL theory has the standard linear theory as its weak-field limit. Expand $\frac{\partial\tU}{\partial\f_{,\m}^a}$ in the field equation (\ref{nuged}) around zero. Let $n>2$ be the lowest order, beyond the Hessian, for which not all the derivatives of $\tU$ vanish at zero\footnote{For example, if we require space-time reflection invariance, $\tU$ is even in $\Gf^a$; so $n\ge 4$; $n=4$ would be generic.}, Write $\f^a=\bar\f^a+\eta^a$, where $\bar\f^a$ is the solution of the linear problem (with the same BC) and expand up to first nonvanishing order $n$:
 $$ q^a\approx\left\{\frac{\partial\tU}{\partial\f^a_{,\m}}(0)
 +\frac{\partial^2\tU}{\partial \f^a_{,\m}\partial \f^b_{,\n}}(0)(\bar\f^b_{,\n}+\eta^b_{,\n})\right\}_{,\m}+$$
\beq +\left\{\frac{1}{(n-1)!} \frac{\partial^n\tU}{\partial \f^a_{,\m}\partial\f^{a_1}_{,{\m_1}}...\partial \f^{a_{n-1}}_{,{\m_{n-1}}}}(0)[(\bar\f^{a_1}_{,{\m_1}}+\eta^{a_1}_{,{\m_1}})
...(\bar\f^{a_{n-1}}_{,{\m_{n-1}}}+\eta^{a_{n-1}}_{,{\m_{n-1}}})]
 \right\}_{,\m},\eeqno{nutceka}
 with repeated $a$ indices summed over.
For the theory to have the linear limit, $\Box\bar\f^a=q^a$, for weak fields, as is assumed, it follows that
$\frac{\partial\tU}{\partial\f^a_{,\m}}(0)=0$, and that
$\frac{\partial^2\tU}{\partial \f^a_{,\m}\partial \f^b_{,\n}}(0)=\d^{ab}\d^{\m\n}$.
Thus, the next order correction $\eta^a$ is gotten as the solution of
 \beq \Box\eta^a=-\frac{1}{(n-1)!} \frac{\partial^n\tU}{\partial \f^a_{,\m}\partial\f^{a_1}_{,{\m_1}}...\partial \f^{a_{n-1}}_{,{\m_{n-1}}}}(0)(\bar\f^{a_1}_{,\m_1}
...\bar\f^{a_{n-1}}_{,{\m_{n-1}}})_{,\m}.\eeqno{supeyu}
In the corresponding PL theory, the solutions of the linear theory are the same. The next order correction is gotten in a similar way:
 \beq \Box\eta^a=\frac{1}{(n-1)!} \frac{\partial^n\tS}{\partial \f^a_{,\m}\partial\f^{a_1}_{,{\m_1}}...\partial \f^{a_{n-1}}_{,{\m_{n-1}}}}(0)(\bar\f^{a_1}_{,\m_1}
...\bar\f^{a_{n-1}}_{,{\m_{n-1}}})_{,\m}.\eeqno{supareq}
So, equivalence of the theories to this order follows if all the $n$th derivatives of $\tS$ and $\tU$ at zero argument are equal in magnitude and opposite in sign.
\par
Since $\tU(x_a)$ and $\tS(y^a)$ are mutual Legendre transforms (I write, for brevity, $x_a$ for the variables $\f^a_{,\m}$ of $\tU$, and $y^a$ for the variables $\psi^a_{,\m}$ of $\tS$) their Hessians are the inverses of each other (at all values of the argument):
 \beq\frac{\partial^2\tU}{\partial x_a\partial x_b}\frac{\partial^2\tS}{\partial y^b\partial y^c}=\d^a_c. \eeqno{gurkla}
So also $\partial^2\tS/\partial y^a\partial y^b(0)=\d_{ab}$ [and $\partial\tS/\partial y^a(0)=0$].
Taking successive derivatives of eq.(\ref{gurkla}) with respect to the $x_a$s at zero argument (and noting that $\partial y^a/\partial x_b(0)=\d^{ba}$), we find, first, that all the derivatives of $\tS$ of order between 3 and $n-1$ also vanish at zero arguments, and that
\beq\frac{\partial^n\tU}{\partial x_{a^1}...\partial x_{a^n}}(0)=-\frac{\partial^n\tS}{\partial x_{a^1}...\partial x_{a^n}}(0), \eeqno{jutakla}
thus confirming that the two theories give the same $\eta^a$.
\{Compare with the condition below eq.(\ref{kipura}), where is is assumed tacitly that $n=4$: when $\tU=\tU[(\Gf)^2]$, the fourth derivative of $\tU$ with respect to component of $\Gf$, at zero, is proportional to $\tU''(0)$.\}

\subsection{Phantom charges}
For theories that have a linear limit--either for weak fields, as in
the systems (i-iii) mentioned in the Introduction, or for strong fields, as in MOND--it is useful to introduce the notion of the ``phantom''  charge density, $q_p(x^\m)$ (or ``phantom'' current density in the electromagnetic case) (PC).
The PC is the charge distribution we have to add to $q$ to make  $\f$, a solution of the linear equation with the same BC.
In other words,
\beq q_p\equiv \Box\f-q.\eeqno{muiop}
If the solution of our theory is unique for given $q(x^\m)$ and BCs, knowledge of $q_p(x^\m)$ is equivalent to knowledge of $\f$.
\par
This concept is useful because it may help us bring our experience with the linear problem to bear on the nonlinear problem, if we have some knowledge of properties of the PC.
\par
For a genuinely NL theory, we cannot know $q_p$ before we solve the full problem. But in PL theories of type (\ref{mudder}),  $q_p$ is known once we have the solution of the linear theory:
 \beq q_p=2(\S'\Gmn\psi_{,\m})_{,\n} -q=q_\f-q. \eeqno{nuiki}
\par
In modified gravity theories such as nonrelativistic MOND, the phantom charge (mass, in this case) represents what we would interpret as ``dark matter'' if we insist that the Poisson equation  governs the gravitational potential $\phi$, when, in fact, it is the NL theory that does.
Much use of the PC has been made in this context (see, e.g., \cite{famaey12} for a review).
\par
Note, importantly, that, unlike $q$, the PC is not an independent quantity that can be dictated at will: given the BCs it is fully determined by $q(x^\m)$.
\par
As an example of some preknowledge of properties of the PC,
consider a static problem in Euclidean space, with a linear, weak-field limit [e.g., of the tree types (i-iii) mentioned above] and assume that $q(\vx)$ is bounded and has a finite total charge. Applying Gauss's theorem to eq.(\ref{muver}) we see that the weak limit is approached at spacial infinity. Thus, $\U'\rar 1/2$ in this limit and so it is seen that
the total phantom charge vanishes: $\int  q_p(\vx)d^3x=0$. This is clearly true also for the PL theory (\ref{mudder}), where, again, applying a Gauss integration over the whole volume gives
$\int  q\_{\f}(\vx)d^3x=\int  q(\vx)d^3x$; so that $\int  q_p(\vx)d^3x=0$.
This is not the case in a theory like MOND where the linear theory is approached in the strong-field limit, not in the weak-field one.
Here, for an isolated mass, the total phantom mass diverges, since the phantom density decreases as $1/r^2$ at infinity.

\section{\label{bi}Electromagnetic vector theories}
The standard Maxwell Lagrangian is:
\beq \L=-\frac{1}{4}\fmn\Fmn+\L_M(A_\m,...), \eeqno{lip}
where the basic DoF are the components of the vector potential $A_\m$, such that $\fmn=A_{\n,\m}-A_{\m,\n}$, and the dependence of $\L_M$ on the other DoF is suppressed. The resulting field equations are
 \beq \Fmn_{,\n}=J^\m,  \eeqno{maxwell}
(plus the identities--the homogeneous Maxwell equation: $F_{(\m\n,\a)}=0$.) where the current $J^\n$ is, as usual, such that $\d I_M=\int J^\n \d A_\n$.
\par
In more general, nonlinear electrodynamics, such as Born-Infeld (BI), we have a Lagrangian of the form
\beq \L=-\tU(\fmn)+\L_M(A_\m,...). \eeqno{buta}
The resulting field equations for $A_\m$ are
\beq  2\left(\frac{\partial\tU}{\partial\fmn}\right)_{,\n}=J^\m.\eeqno{born}
\par
Lorentz invariance dictates that the Lagrangian depends on $\fmn$ through its invariants. In four dimensions (which I assume all along for concreteness) these are the two invariants
\beq \pf=\frac{1}{4}\fmn\Fmn=\oot(B^2-E^2),~~~~\qf=\frac{1}{8}\eps^{\a\b\m\n}
F_{\a\b}\fmn=-\oot\vE\cdot\vB, \eeqno{lirom}
where $\vE$ and $\vB$ are the electric and magnetic fields\footnote{I work with a $(-1,1,1,1)$ signature, and $c=1$, $\eps_{\a\b\m\n}$ is the totally antisymmetric tensor; $\eps_{0123}=1$; so, $\eps^{0123}=-1$.}.
So write
\beq \L=-\U(\pf,\qf)+\L_M(A_\m,...). \eeqno{butama}

For example, in the standard BI theory
 \beq \tU=E_0^2(-\|\emn+\fmn/E_0\|)^{1/2}, \eeqno{mature}
up to a constant, which is immaterial here, as I assume flat space-time.
This can be written in four dimensions as
 \beq \U =E_0^2\left(1+\frac{2\pf}{E_0^2}-\frac{4\qf^2}{E_0^4}\right)^{1/2}
 =E_0^2\left[1-\frac{E^2-B^2}{E_0^2}-\frac{(\vE\cdot\vB)^2}{E_0^4}\right]^{1/2}. \eeqno{mureda}

In terms of $\U(\pf,\qf)$ we can write the field equations as
 \beq (\U_p\Fmn+\U_q\tFmn)_{,\n}=J^\m,  \eeqno{lumita}
where $\U_p$ and $\U_q$ are the partial derivatives of $\U$, and $\tFmn=(1/2)\eps^{\m\n\a\b}F_{\a\b}$. (The homogeneous identities can be written as $\tFmn_{,\n}=0$.)

For the theory to tend to Maxwell's in the limit of weak fields we
need $\U_p^0\equiv\U_p(0,0)=1$, and  $\U_q^0\equiv\U_q(0,0)=0$.
We see that for the BI theory, we have, in addition,  $\U_q(p,0)=0$, for all $p$. This is more generally the case in theories with time-reversal invariance, since $q$ changes sign under time reversal. I will assume this in what follows.
\subsection{Practically linear kindred}
Now, consider a class of PL electrodynamic theories
that involve, in addition to $A_\m$, an auxiliary vector field $B_\m$, with the corresponding $\qmn=B_{\n,\m}-B_{\m,\n}$, and having the Lagrangian
 \beq \L=-\oot\fmn\Qmn+ \tS(\qmn)+\L_M(A_\m,...), \eeqno{kul}
where $\L_M$ is the standard matter Lagrangian; it depends in the standard way on $A_\m$ (but not $B_\m$), and on the other matter DoF.
Variation over $A_\m$ now gives
\beq \Qmn_{,\n}=J^\m. \eeqno{niy}
Namely, $B_\m$ is a Maxwellian EM vector field for the given current distribution.
Variation over $B_\m$ gives
\beq \Fmn_{,\n}=\J^\m\equiv 2\left(\frac{\partial \tS}{\partial\qmn}\right)_{,\n}. \eeqno{nulup}
So the EM field $A_\m$ is also a Maxwell field but for the current $\J^\m$.  This current is identically conserved, because  $\Pmn\equiv 2\partial \tS/\partial\qmn$ is antisymmetric in the indices.
$\Pmn$ is an algebraic expression of $Q_{\a\b}$, the Maxwellian field of the problem.
The theory has the double gauge invariance\footnote{
We can write the result above in terms of the Hodge decomposition of the 2-form $P=P_{\m\n}dx\^\m\wedge dx\^\n$:
If we decompose (the decomposition is unique when appropriate BC are imposed, effectively compactifying the underlying space)
 \beq P=dA+ \ddag B^{(3)}+h,  \eeqno{musht}
where, $A=A_\m dx^\m$ is a vector potential, $B^{(3)}$ is some 3-form, and $h$ is harmonic (a vacuum solution). Then
\beq F=\fmn dx\^\m\wedge dx\^\n=dA+ h.  \eeqno{mushtat}
It satisfies $dF=0$, the homogeneous equation (the Bianchi identities), since harmonic forms are closed (and co-closed).
It, clearly, also satisfies eq.(\ref{nulup}), which can be written as $\ddag(F-P)=0$; the homogeneous identities $dF=0$ are also satisfied}.
\par
In four dimensions it is convenient to write the $\tS$ as a function of the invariants
 \beq \L=-\oot\fmn\Qmn+ \S(\pq,\qq)+\L_M(A_\m,...). \eeqno{kular}
The field equations (\ref{nulup}) are then written as
 \beq \Fmn\_{,\n}=(\S_p\Qmn+\S_q\tQmn)\_{,\n}. \eeqno{gutrab}
Here, the right-hand side is given once the solution $\qmn$ of the Maxwell equations of the problem is known.
\par
It is sometimes useful to write the theory in terms of $\hmn\equiv\fmn-\qmn$ instead of $\fmn$:
 \beq \Hmn\_{,\n}=[(\S_p-1)\Qmn+\S_q\tQmn]\_{,\n}. \eeqno{gutkip}
To insure the Maxwellian weak-field limit we have $\S_p^0=\S_p(0,0)=1$, and $\S_q^0=\S_q(0,0)=0$. We can then write to lowest order in weak fields
\beq \L_0=-\frac{1}{4}(\fmn\Fmn-H_{\m\n}H^{\m\n})+\L_M(A_\m,\psi), \eeqno{kumneq}
We see that, again, $A_\m$ has the standard Maxwell action and couples to currents in the standard way, while $E_\m\equiv A_\m-B_\m$ decouples altogether. Note, however, that, as in the scalar case, this latter, auxiliary field, has the ``wrong'' (ghost-like) sign of its kinetic action.

\subsection{Similarity conditions}
What choices of $\tS$ give a PL theory that is equivalent to the NL theory for 1-D configurations, and equivalence to second order in the field gradients, for any configuration.
\subsubsection{Equivalence for static 1-D problems}
For a static, 1-D configuration, currents depend only on one space coordinate, $x_m$, in an orthogonal coordinate system (e.g., a time-independent, spherically symmetric charge distribution, or a current density distribution in an infinite cylindrical wire).
\par
Staticity and current conservation imply $\div\vJ=0$. Also, $\div\vB=0$. Applying Gauss theorem to surfaces of constant $x_m$ shows that $J^m=0$, $B_m=0$. These apply in both theories, to both $J^\m$ and $\J^\m$, and to all the magnetic fields (i.e., those in both $\qmn$ and $\fmn$). Also, the electric fields, being gradients, can have only an $m$ component; thus $\qf=\qq=0$.
\par
If the Lagrangian is stationary in $q$ at $q=0$--as is the case in the BI theory, and more generally in theories in which $q$ appears quadratically in the Lagrangian (e.g., forced by time-reversal invariance)--which I assume, we have  $\U_q(\pf,0)=0$. So, the field equations (\ref{lumita}) can now be written (writing the current in terms of the Maxwellian solution: $J^\m=Q^{\m \n}\_{,n}$)
\beq [\U_p(\pf,0)F^{\m k}-Q^{\m k}]_{,k}=0.  \eeqno{lumza}
Similarly, for the PL theory the field equations (\ref{gutrab}) read
 \beq [F^{\m k}-\S_p(\pq,0)Q^{\m k}]\_{,k}=0. \eeqno{gutpil}
\par
Applying a Gauss theorem for the time component, and Ampere theorem for the space components show that in both theories, the expression in parentheses vanish for our static 1-D configurations:
\beq \U_p(\pf,0)F^{\m k}=Q^{\m k},  \eeqno{lumzama}
for the nonlinear theory, and
 \beq F^{\m k}=\S_p(\pq,0)Q^{\m k}, \eeqno{gutpilma}
for the PL theory.
[The fields $Q^{\m k}$ are the same in both theories.] So, clearly, 1-D equivalence of the theories is tantamount to
 \beq \bU'(\pf)\bS'(\pq)=1,  \eeqno{kurpola}
for the values of $\pf$ and $\pq$ that correspond to each other (in either theory), and where I defined $\bU(\pf)\equiv \U(\pf,0),~\bS(\pq)\equiv \S(\pq,0)$.
\par
To find the relation between $\pf$ and $\pq$, contract each side of equations (\ref{lumzama})(\ref{gutpilma}) with itself, to get
 \beq [\bU'(\pf)]^2\pf=\pq, ~~~~~~~~[\bS'(p\_Q)]^2p\_Q=\pf, \eeqno{mioles}
respectively.
\par
To recap, given $\U$, if we choose $\S$ so that eq.(\ref{kurpola}) is satisfied, for $\pf$ and $\pq$ related by either of equations (\ref{mioles}), we get a PL theory with 1-D equivalence to the NL theory governed by $\U$. Thus, $\S(\pq,0)$ is determined uniquely (up to an additive constant), but not, of course, $\S(\pq,\qq)$
\par
For BI, where $\bU(x)=E_0^2(1+2x/E_0^2)^{1/2}$, one gets from the above requirement of similarity, $\bS(y)=-E_0^2(1-2y/E_0^2)^{1/2}$ (up to a constant). So we get for the BI case
$\bS(y)=-\bU(-y)$.

\subsubsection{Second-order equivalence}
Another way to constrain the dependence of $\S$ on $p,~q$ is to require that the two theories coincide for a general problem up to next to lowest order in the fields (if we require in the first place that both coincide with Maxwell's linear theory to lowest order).
\par
Start with eq.(\ref{lumita}), where we write $\fmn=\qmn+\hmn$, with $\qmn$ the Maxwellian solution. Subtracting the zeroth, Maxwellian order, we are left with
\beq -\Hmn_{,\n}=\U^0_{pp}(\pq\Qmn)_{,\n}+\U^0_{pq}(\qq\Qmn+\pq\tQmn)_{,\n}+
\U^0_{qq}(\qq\tQmn)_{,\n},  \eeqno{lumbar}
where the lowest order, Maxwellian solution is substituted everywhere in the right-hand side. Similarly, in the PL theory we get to this order
\beq \Hmn_{,\n}=\S^0_{pp}(\pq\Qmn)_{,\n}+\S^0_{pq}(\qq\Qmn+\pq\tQmn)_{,\n}+
\S^0_{qq}(\qq\tQmn)_{,\n}.  \eeqno{lumsaf}
The two theories are then equivalent to this order in the fields if, in addition to $\U^0_p=\S^0_p$ and $\U^0_q=\S^0_q=0$, we have $\U^0_{ij}=-\S^0_{ij}$, similar to the conditions (\ref{jutakla}) in the multi-scalar case.

\subsubsection{The Legendre connection}
Unlike the single-scalar case, where the requirement of equivalence for static, 1-D configurations, determines $\S(y)$ given $\U(z)$, here $\S(\pf,0)$ is determined, given $\U(\pq,0)$, but not its dependence on $\qf$. Also, unlike the single-scalar case, this condition, alone, does not insure equivalence of the theories to next order in the fields; for this we further require conditions on the second derivatives of $\S$ at zero arguments.
\par
Taking a cue from the scalar case, we choose
$\tS$ to be the Legendre transform of $\tU$ in the six field variables\footnote{Not in $p$ and $q$, and not, e.g., with respect to $\vE$ alone, which would give the Hamiltonian of the theory.} $\fmn$, or in $\vE$ and $\vB$.
It is then seen that if we opt for a Palatini variation, namely, we consider the antisymmetric $\qmn$ a basic DoF, without forcing it to be a curl, we get the genuinely NL theory (\ref{buta}-\ref{born}). If, however, we do constrain $\qmn$ to be the curl of a vector $B_\m$, we get the PL kindred of this theory, as discussed above. This PL theory automatically satisfies the above two similarity requirements:
One expresses the Hessians of $\tU$ and $\tS$ in terms of the partial derivatives of $\U$ and $\S$ in $p$ and $q$. Then one uses the fact that the Legendre connection implies that Hessians are mutual inverses, for corresponding values of their variables, to show that: a. $\bU'(\pf)\bS'(\pq)=1$ (as well as various other relations), and b. that at zero-fields $\U^0_{ij}=-\S^0_{ij}$.
\par
For the BI case, start with $\tU(\vE,\vB)$ given in eq.(\ref{mureda});
define as usual
 \beq\vD\equiv \frac{\partial \L}{\partial\vE}=-\frac{\partial \tU}{\partial\vE}, ~~~~~\vH\equiv -\frac{\partial \L}{\partial\vB}=\frac{\partial \tU}{\partial\vB}.
 \eeqno{legaga}
Then, the Legendre transform of $\tU(\vE,\vB)$ is
 \beq \tS(\vD,\vH)=-\vD\cdot\vE+\vH\cdot\vB-\tU=
 -E_0^2\left(1-\frac{2\pq}{E_0^2}-\frac{4\qq^2}{E_0^4}\right)^{1/2}
 =-E_0^2\left[1+\frac{D^2-H^2}{E_0^2}-\frac{(\vD\cdot\vH)^2}{E_0^4}\right]^{1/2}. \eeqno{murpsa}
\par
It is easily checked directly that with this choice of $\S$, both the above conditions for equivalence in 1-D configurations, and equivalence to second lowest order in the fields for any configuration, are satisfied. As said above, these conditions do not determine $\S$ uniquely, as the first concerns only its dependence on $\pq$ for $\qq=0$, and the second constrains only some derivatives at zero fields. But this choice of $\S$ might have additional advantages, which I have not yet pinpointed.
\par
The BI Lagrangian is special in that it has the same form as its Legendre transform, only with the opposite sign of $E_0^2$ [compare with eq.(\ref{mureda})]. In other words $\tS(\qmn,E_0^2)=\tU(\qmn,-E_0^2)$. Or, in determinant form,
\beq \tS=-E_0^2(-\|\emn+i\qmn/E_0\|)^{1/2}, \eeqno{matanela}

\section{\label{rk}A toy example: special-relativistic kinematics}
Consider now the construction of an analog PL theory for the even simpler problem of point particles in Minkowski space-time, with world lines $x^\m_k(\tau)$ ($k$ is the particle index).
The action is
 \beq I=-\sum_k m_k\int d\tau_k +I_{int},  \eeqno{kupile}
$I_{int}$ being the interaction action, which depends on the $x^\m_k(\tau)$.
Pick some Lorentz frame in which the particle trajectories in space are $\vx_k(t)$, for which the free particle action is $-\sum_k m_k\int dt(1-{\vv}_k^2)^{1/2}$ (with $\vv_k=\dot\vx_k$). The equations of motion are $m_k d[{(1-\vv_k^2)^{-1/2}\vv}_k]/dt=\vF_k(\vx_1,\vx_2,...)=\d I_{int}/\d \vx_k$.
In the PL analog we add auxiliary degree of freedoms $\vy_k(t)$ and the action is
\beq I=\sum_k m_k\int dt[\dot{\vx}_k\cdot\dot{\vy}_k-\oot \l(\dot y_k^2) ]+I_{int},  \eeqno{kupisa}
where the interaction depends only on $\vx_k$.
Varying over $\vx_k$ and $\vy$ respectively
gives
 \beq m_k\ddot{\vy}_k=\vF_k, ~~~~~~\ddot{\vx}_k=d[\l'(\dot y_k^2)  \dot{\vy}_k]/dt.
\eeqno{niure}
In other words $\vy_k(t)$
is the solution of Newton's equations for the same run of the forces on the particles as in the relativistic problem\footnote{This means that to get $\vy_k(t)$ we first have to know the $\vx_k(t)$, calculate for these the time runs of the forces $\vF_k$, and calculate the $\vy_k(t)$ as the Newtonian trajectories for these forces.
Alternatively, we could have considered a problem in which the forces $\vF_k(t)$ are dictated in some Lorentz frame, instead of the interactions. Then $\vy_k(t)$ are the Newtonian trajectories for these forces, and $\vx_k(t)$ are the special-relativistic ones.},
and once it is solved for and substituted in the right-hand side of the second equation, we have another Newtonian equation to solve.
\par
To lowest order in the velocities, the free Lagrangian (for a unit mass, and dropping the particle index from now on) is $L_K\approx \dot{\vx}^2/2-(\dot {\vx}-\dot{\vy})^2/2$.
We see, again, that the difference degree of freedom has the ``ghost-like'' sign of the kinetic term, but this decouples altogether from other DoFs. Is this a bad sign for the theory?
\par
The second eq.(\ref{niure}) integrates to $\dot{\vx}=\l' \dot{\vy}$ (with the appropriate initial conditions). We see that the two velocities are always parallel, and squaring this relation $\dot x^2=[\l'(\dot y^2)]^2\dot y^2$, gives an algebraic relation between their magnitudes; so $\dot{\vy}$ can be algebraically expressed in terms of $\dot{\vx}$. It is seen that if we take $\l(z)=2(1+z)^{1/2}$ [$-\l/2$ is the Legendre transform of $(1-\vv_k^2)^{1/2}$], and plug the expression of $\dot y^2$ in terms of $\dot x^2$ back in the action, eliminating the dependence on $\vy_k$, we get the standard Lorentz Lagrangian (\ref{kupile}).
So, in the above chosen frame, the action
 \beq I=\sum_k m_k \int dt[\dot{\vx}_k\cdot\dot{\vy}_k-(1+\dot y_k^2)^{1/2} ]+I_{int},  \eeqno{kupapa}
gives an equivalent theory. This action is not Lorentz-invariant, but gives invariant equations of motion for the $x^\m_k(\tau)$.
\par
For the kinetic Hamiltonian,  $H_K=\sum \dot{\vec\xi}_i\cdot\vP_i-L_K$, where $\vP_i=\partial L/\partial \dot{\vec\xi}_i$ [$\vec \xi\equiv(\vx,\vy$)], we find $H_K=\dot{\vx}\cdot \dot{\vy}-\l'\dot y^2+\l/2$. For solutions of the equations of motion the first two terms cancel and we are left with $H_K=\l/2=(1+\dot y^2)^{1/2}$, which is always positive.
We can also express it in terms of $\dot x^2$, through the algebraic relation, which gives the standard special-relativistic energy $H_K=(1-\dot x^2)^{-1/2}$. So despite the alarming appearance of ghost-like terms in the linear limit, the theory is stable and otherwise healthy.
The two theories have in fact the same solutions under the same initial conditions.
\par
The identity of the two theories follows immediately from the fact that the $t$ is the only independent variable of the degrees of freedom; so the constraint that would differentiate between the theories is not a real constraint.
The PL construction is not of practical use in this easily integrable case, but it is a useful heuristic example.

\section{\label{electrostatics}Example: fields and forces in nonlinear electrostatics}
I now look more closely at the specific example of BI electrostatics, to highlight some of the similarities and dissimilarities between this theory and its kindred PL theory. Some related issues in the context of the BI theory were discussed in \cite{gibbons98}.
\par
Consider an electrostatic configuration made of a charge distribution $\r(\vr)$ in 3-D Euclidean space. The action for a general, genuinely NL  theory of electrostatics, with one potential, to which the general Lagrangian (\ref{buta}) leads, can be written as
  \beq I= \int \{ -u[\gfs]-\r\f\}~d^3r ,\eeqno{pushalk}
where $\f$ is the electrostatic potential ($\vE=-\gf$). [I have changed the notation a little: here it is convenient to use $x\equiv \gfs$ and $y\equiv (\gps)^2$ as variables, and the Lagrangian functions for the two types of theory will be denoted $u(x)$ and $s(y)$.
I also choose the arbitrary additive constant in $u$ and $s$ so that $u(0)=s(0)=0$.]
The field equation is
\beq 2\div(u'\gf)=2u'\left(\d_{ij}+2\hat u' \frac{\f\_{,i}\f\_{,j}}{|\gf|^2}\right)\f_{,i,j}=\H^u_{ij}\f_{,i,j}=\r,  \eeqno{muipo}
where, $\H^u_{ij}$ is the Hessian of $u$ with respect to the variables $\f_{,i}$, and $\hat u'=xu''(x)/u'(x)$ is the logarithmic derivative of $u'$.
\par
If the theory has the standard, linear, weak-field limit, as I assume in what follows\footnote{Since we are dealing with electrostatics, $u$ is negative, so as to give repulsion for like charges. In NL theories of gravity, such as MOND, $u$ is positive.}, $u(x)\approx -x/2$ for $x\rar 0$, and thus $\hat u'(0)=0$. Ellipticity of the field equation, which I require, is tantamount to $\H^u_{ij}$ being regular. Since one of its eigenvalues is $\propto 1+2\hat u'$, which must not vanish, we have that $\hat u'(x)>-1/2$ for all $x$. Another useful inequality follows from this, with our normalization $u(0)=0$: $\hat u\equiv xu'/u> 1/2$. To see this consider the function $w(x)=xu'(x)-u(x)/2=u(\hat u-1/2)$. We have $w(0)=0$, and $w'(x)< 0$, by virtue of the above inequality for $\hat u'$, and the fact that $u'<0$. So $w(x)\le 0$, and vanishes only for $x=0$. Thus $w/u>0$ for $x>0$ (since $u<0$) leading to the required inequality.
\par
The corresponding PL action, gotten from the Lagrangian (\ref{kul}), is
\beq I=\int \{\gf\cdot\gps +s[(\gps)^2]-\r\f\} d^3r \eeqno{mudderlk}
Variation on $\f$ and $\psi$, respectively, gives
 \beq \Delta\psi=-\r,~~~~~~\Delta\f+2\div(s'\gps)=
 \Delta\f+2s'\left(\d_{ij}+2\hat s' \frac{\psi\_{,i}\psi\_{,j}}{|\gps|^2}\right)\psi_{,i,j}
 =\Delta\f+\H^s_{ij}\psi_{,i,j}=0,  \eeqno{megnaio}
where, $\H^s_{ij}$ is the Hessian of $s$, and $\hat s'=ys''(y)/s'(y)$.
\par
We saw that in this, single-potential case, the requirement of equivalence for 1-D configurations fixes $s$ uniquely as the Legendre transform of $u$ (in the variables $\gf,~\gps$).
Thus, $\H^s_{ij}$ and $\H^u_{ij}$ are mutual inverses at the corresponding values of the variables, and the inequalities $\hat s'> -1/2$, $\hat s=ys'/s> 1/2$  apply for $s$, as for $u$.
\par
Unlike the field equations (\ref{muipo}), which generally require numerical solution, the solution of the PL theory can be written in closed form as space integrals:
For example, for a system of point charges $q_i$ at $\vr_i$
 \beq \gps(\vr)=\frac{1}{4\pi}\sum_i \frac{q_i(\vr_i-\vr)}{|\vr_i-\vr|^3}, \eeqno{milfa}
and
 \beq\f(\vr)=\frac{1}{4\pi}\int  \frac{\div\vA(\vr')}{|\vr'-\vr|}d^3r', \eeqno{mlawerda}
 where, $\vA\equiv 2s'\gps$.
Integrating by parts gives other useful expressions (for theories with standard weak-field limit, as I assume here, the surface integral at infinity vanishes):
 \beq\f(\vr)=\frac{1}{4\pi}\int  \frac{(\vr'-\vr)\cdot \vA(\vr')}{|\vr'-\vr|^3}d^3r'
 =\frac{1}{4\pi}\int  \frac{\hat\vr\cdot\vA(\vr+\hat\vr)}{|\hat\vr|^3}d^3\hat r, \eeqno{klowerda}
 \beq\gf(\vr)=\frac{1}{4\pi}\int \div\vA(\vr')\frac{\vr'-\vr}{|\vr'-\vr|^3}d^3r' , \eeqno{mupda}
and after integration by parts
 \beq\gf(\vr)=-\frac{1}{4\pi}\int \frac{ d^3r'}{|\vr'-\vr|^3}
 \left[I-3\frac{(\vr'-\vr)\otimes(\vr'-\vr)}{|\vr'-\vr|^2}\right]\cdot\vA(\vr') , \eeqno{muparuy}
 where $I$ is the unit matrix. Changing variables:
  \beq\gf(\vr)=-\frac{1}{4\pi}\int \frac{d^3\hat r}{|\hat\vr|^3} \left[I-3\frac{\hat\vr\otimes\hat\vr}{|\hat\vr|^2}\right]\cdot\vA(\vr+\hat\vr). \eeqno{muperioy}
This converges at infinity because $\vA$ vanishes there, and at $\hat\vr=0$, because the angular integrals for constant $\vA$ are of spherical harmonics, and vanish.

\par
For BI electrostatics,
 \beq u(x)=E_0^2[(1-x/E_0^2)^{1/2}-1], ~~~~~~s(y)=E_0^2[1-(1+y/E_0^2)^{1/2}]; \eeqno{opopka}
so $\vA=-\gps/(1+|\gps|^2)^{1/2}$.
Interestingly, $s$ is then the Lagrangian for the volume extremization problem. Taking from now on $E_0=1$, we have
here $\hat s'(y)=-y/[2(1+y)]$.
\par
The phantom densities in the two theories are
 \beq \r_p=\r[(1-E^2)^{1/2}-1]-\frac{E}{1-E^2}\vE\cdot\grad |\vE|, \eeqno{mitfer}
in BI, while in the PL theory it is
 \beq \r_p=\r[\frac{1}{(1+E\_C^2)^{1/2}}-1]
 -\frac{E\_C}{(1+E\_C^2)^{3/2}}\vE\_C\cdot\grad |\vE\_C|, \eeqno{mitliq}
where $\vE\_C$ is the Coulomb field.
The total phantom charge vanishes in both theories.
\par
The two theories are equivalent for 1-D configurations, in which case expressions (\ref{mitfer}) and (\ref{mitliq}) are seen to be the same.
\subsection{Integral relations for the PL theory}
There are several useful integral relations. First note that integrating the second of eq.(\ref{megnaio}) for $\f$ over any volume bounded by a surface $\Sigma$, we have
 \beq \int\_\Sigma\gf\cdot \dvs=-\int\_\Sigma 2s'\gps\cdot \dvs, \eeqno{liraf}
on the surface.
Now, multiply that equation by $\psi$, and integrate over a volume $V_\psi$, bounded by an equipotential surface, $\Sigma$, of $\psi$. Integrating by parts, the surface integrals are equal, by relation (\ref{liraf}), and cancel, and we get
 \beq \int_{V_\psi} \gf\cdot\gps=-2\int_{V_\psi} s'|\gps|^2d^3r.  \eeqno{virial}
A useful corollary is that for any such volume, the integral of the free-field Lagrangian density  in expression (\ref{mudderlk}) is
 \beq I_f(V)=\int_{V_\psi} (\gf\cdot\gps +s) d^3r =
 \int_{V_\psi} (s-2s'|\gps|^2) d^3r=\int_{V_\psi} s(1-2\hat s) d^3r.\eeqno{beinam}
This integral is nonnegative, and vanishes only if $\gps=0$ everywhere in the volume, since $\hat s(z)>1/2$, as shown above, and $s\le 0$ with equality only at $z=0$. This applies, in particular when integrating over the whole space.
\par
Now multiply the field equation by $\f$, and integrating in a volume, ${V_\f}$, bounded by an equipotential of $\f$. We then get in the same way
 \beq \int_{V_\f} \gfs=-\int_{V_\f} 2s'\gf\cdot\gps.  \eeqno{merfac}
For ${V_\f}$ the whole space, both equalities apply.

\subsection{Properties of the field}
There is much known about the solutions of elliptic equation of the type (\ref{muipo}) (see, e.g., \cite{gilbarg77}). Paradoxically, even though the PL kindred is simpler to solve, I am not aware of discussions of the analog properties for it. I now discuss briefly several of these properties, where I mainly pose questions and suggest some insights pertaining to possible answers.
\subsubsection{Extrema of the potential}
It is well known that solutions of eq.(\ref{muipo}) cannot attain extrema in vacuum, except on boundaries (this is known as
a maximum principle \cite{gilbarg77}). This means, for example, that we cannot suspend a test charge in an electrostatic field, outside source charges\footnote{It was shown in \cite{milgrom98} that it is possible to suspend a non-test charge, or a rigid body of test charges, in vacuum in certain configurations.}; is this also true for the `physical' potential, $\f$, in the PL analog? (It is true for $\psi$, of course.)
\par
I was not able to find an answer for this in the mathematical literature. The potential $\f$ satisfies the Poisson equation with $\r+\r_p$ as source, but while $\r$ may be localized on a finite support, $\r_p$ is not, in general. We see from eq.(\ref{mitliq}) that this density does vanish where $\r=\vE\_C\cdot\grad |\vE\_C|=0$. For example, this holds at all the critical points of the Coulomb potential of a system of point charges (where $\vE\_c=0$). So, for example, $\f$ does not attain an extremum at symmetry points (such as the midpoint between equal charges). It is also true in 1-D configurations.

\subsubsection{Boundedness of the electric field}

In BI, the electric field strength $E=|\gf|$ is bounded from above by 1 ($E_0$); is this also the case in the PL theory? I have not been able to answer this question in general.
\par
In the 1-D configurations this is clearly the case, since then $\vE=\vA\equiv\vE\_C(1+ \vE\_C^2)^{-1/2}$. So, $E$ is indeed bound, and   $E\rar 1$ when approaching source singularities, such as point charges. But is this true always? In general we only know that $\div\vE=\div\vA$, from which we can derive average upper limits for $E$. For example, eq.(\ref{liraf}) tells us that for any closed surface \beq\int\gf\cdot d\s =\int\vA\cdot \dvs\le\int|\vA|d\s\le \int d\s. \eeqno{astralii}
If the surface is an equipotential of $\f$ (on which $\gf\cdot\dvs=Ed\s$) we have
\beq \int E d\s \le \int d\s.  \eeqno{astrali}
In other words, the area-weighted average of $E$ on any closed, $\f$ equipotential surface does not exceed 1.
\par
Also, from eq.(\ref{merfac}), we have for a volume within such a surface
\beq \int_{V\_\f}E^2=\int_{V\_\f}\vE\cdot\vA\le \int_{V\_\f}E|\vA|\le \int_{V\_\f}E. \eeqno{astraliii}
This means $ \int_{V\_\f}E(E-|\vA|)\le 0$; so $E$ cannot exceed $|\vA|$ everywhere in the volume.

\subsection{Forces on bodies}
Consider the force, $\vF\_V$, acting on a subsystem of $\r(r)$ made of all the charge within some sub-volume $V$. Some relevant results were derived in \cite{milgrom02} for theories governed by NL actions of type (\ref{pushalk}), and in \cite{milgrom10} for theories of type (\ref{mudderlk}).
$\vF\_V$ is writable as an integral over any closed surface, $\Sigma$, that surrounds all the charges in $V$, and excludes all others:
\beq\vF\_V=\int_\Sigma \textbf{P}\cdot \dvs,  \eeqno{muipola}
where $\textbf{P}$ is the stress tensor defined as the functional derivative of the free-field action with respect to the background metric, and $\dvs$ points outward of $V$. More specifically,
\beq \d I_{fields}\equiv \oot\int g^{1/2}\textrm{P}_{ij}\d g^{ij} \eeqno{mipla}
(To identify $\textrm{P}^{ij}$ we write the action on a curved background, after which we can specialize back to a flat background).
For the genuinely NL theory (\ref{pushalk}) this gives:
\beq \vF\_V=\int_\Sigma u \dvs-2u'\gf(\gf\cdot \dvs).\eeqno{forcei}
[This is correct provided $u(0)=0$, otherwise $u-u(0)$ appears instead of $u$.]
For the PL theory (\ref{mudderlk}) we get [again, provided $s(0)=0$]:
\beq \vF\_V=\int_\Sigma -(s+\gf\cdot\gps) \dvs+2s'\gps(\gps\cdot \dvs)+\gps(\gf\cdot \dvs)+\gf(\gps\cdot \dvs).\eeqno{forceii}
For example, to calculate the force between two equal charges we can choose $\Sigma$ as the symmetry plane completed by an hemisphere at infinity, on which the integral vanishes. From symmetry, $\gps\cdot \dvs=\gf\cdot \dvs=0$, and $\gf\parallel\gps$ on the midplane, and we have for the two theories respectively
\beq \vF\_V=\int_\Sigma u \dvs,~~~~~~\vF\_V=-\int_\Sigma (s+|\gf||\gps|) \dvs .\eeqno{forceiv}
\par
To calculate the force in the nonlinear theory we  first need to solve the field equation (\ref{muipo}) given $\r$ and boundary conditions ($\gf=0$) at infinity. We then use the result in the integral (\ref{forcei}). The calculation in the PL theory is more straightforward,  as we have a closed form for
the integrand in expression (\ref{forceii}).
\subsubsection{Force on a spherical charge by a weak charge distribution}
Consider a system made of an arbitrary 1-D  charge distribution $\r\_S$ (e.g., spherical; e.g., a point charge), and an arbitrary charge distribution $\r$ that is so weak that we can treat it as a test-charge distribution. Then, the force on $\r$ (by momentum conservation also minus the force on $\r\_S$) is easily calculated in both theories (e.g., \cite{milgrom98}). This force is simply $\int \r\vE\_S d^3r$, where $\vE\_S$ is the electric field produced by $\r\_S$ alone. Since $\r\_S$ is 1-D, $\vE\_S$ is easily calculated, and furthermore, it is the same in the two theories. For example, it can be shown, based on this, that a point charge $q\_S$ can be suspended at the center of a cube, at the corner of which we have charges $q$ of opposite sign, and $|q|\ll|q\_S|$.
\subsubsection{Attraction and repulsion between bodies of uniform charge sign}
In \cite{milgrom02} I formulated a push-pull conjecture pertaining to the question of attraction and/or repulsion between bodies each made of a charge distribution of a uniform sign. In the present context: Suppose we have two parallel planes, say parallel to the $x-y$ plane, with a charge distribution $\r_1\ge 0$ between the planes, $\r_2\ge 0$ to the right of the two planes, and $\r_3\le 0$ to their left. Then, in a theory like BI, it is was conjectured that the force on $\r_1$ is always to the left.
This conjecture implies, e.g., that bodies of uniform-sign charge separated by a plane always repel each other if they have the same sign (namely the force on each of the bodies points to the other side of any separating plane), and attract for opposite signs. The general conjecture is trivial to prove for the linear, Coulomb electrostatics. But I was able to prove only special cases of this conjecture for BI electrostatics (for example the case where $\r_1$ is spherical and monotonically decreasing from its center out).
There is an even more elementary result that holds for BI: If $\r\ge0~(\le 0)$ are all the charges in space, and $C$ is the convex closure of the support of $\r$ ($C$ is the smallest convex volume containing all points where $\r\not =0$), then, at any point outside $C$ the field E points away from (into) $C$. This I proved in \cite{milgrom02} using a comparison principle for an equation like eq.(\ref{pushalk}).
\par
For the PL kindred I was not able to prove even these special cases.
Of course, if one of the three bodies is 1-D, and the other two are test bodies, the conjecture is easily seen to be correct in the PL theory.

\subsubsection{The two-body force}
One of the interesting problems in NL electrostatics is the calculation of the force between two point charges. On dimensional grounds, the force between charges $q_1$ and $q_2$, a distance $\ell$ apart, can be written as (reinstating $E_0$)
 \beq F(q_1,q_2,\ell)=\frac{q_1q_2}{4\pi \ell^2}f(\lambda,\z), \eeqno{byvat}
(positive for repulsion) where the dimensionless variables are $\lambda\equiv (|q_1|+|q_2|)/4\pi E_0 \ell^2$, and the charge ratio $\z\equiv q_1/q_2$ ($|\z|\le 1$).
It was proven in \cite{milgrom02} that, with the choice of the sign of the free-field action in expression (\ref{pushalk}), point charges of the same (opposite) sign repel (attract) each other so $f\ge 0$ (the opposite is true when the sign of the action is inverted, as in MOND gravity).
The weak-field limit applies for $\lambda\rar 0$, where we have for both theories\footnote{The fact that the field is strong near the point charges is immaterial. When $\l\ll 1$ we can choose the integration surface in eqs.(\ref{forcei})(\ref{forceii}) where the field is weak, and the force attains its weak-field limit (this needs more careful showing, but is correct).} $f(0,\z)=1$. The fact that the two theories are the same to second order tells us that $f$ for the two theories are the same also to first order in $\lambda$. The limit $\z\rar 0$ corresponds to $q_1$ being a test charge in the spherical field of $q_2$, which is known analytically. Thus, the two theories coincide and give:
$f(\lambda,0)=(1+\lambda^2)^{-1/2}$.
For the limit $\lambda\rar \infty$ both theories give a constant force, and we can write then $f(\lambda\rar\infty,\z)\rar \lambda^{-1}\hat f(\z)$.
\par
In \ref{appa} , I calculate $\hat f(\z)$ analytically, in BI electrostatics, for charges of the same sign, for any dimension. For 3-D, I find a repulsive force
\beq F\rar \frac{E_0q_1q_2}{|q_1+q_2|}, \eeqno{aareta}
that is, $\hat f(\z)=1$.
\par
In two dimensions, or for parallel lines of uniform charge (where the charges and the force are per unit length), I get in this limit
 \beq F(q_1,q_2,\ell\rar 0)\rar \frac{E_0|q_1+q_2|}{\pi}
 sin\left(\pi\frac{q_1}{q_1+q_2}\right). \eeqno{hutat}
This is the same as the result in \cite{pryce35}, obtained using a method that applies only in 2-D.
\par
For the PL theory I have not yet been able to derive similar results, as the specific method used in the BI case does not directly apply to it.

\section{\label{discussion}Discussion}
I have presented a class of theories that on one hand describe nonlinear physics, but which require solving only linear differential equations. There are examples where such theories, when properly constructed, serve as very useful approximations for genuinely NL physical systems. However, much still remains to be checked if these theories are to stand by themselves--not only as approximations or heuristic tools--e.g., as theories of gravity, generalizing standard gravity--as in the MOND paradigm--or as theories of electromagnetism.
\par
We saw, for example, that ghosts appear in the weak-field limit of all the versions of the PL theories that have a linear weak-field limit.
But it has to be checked how deleterious they are. In the linear approximation these ghosts decouple altogether and do not affect physics; and it is not clear to what extent they survive nonperturbatively. For example,
in the PL version of special-relativistic kinematics, where (decoupled) ghosts do appear in the linear limit, they seem to be harmless as the theory in full is equivalent to special relativity, and as we saw its Hamiltonian is always positive. We also saw that with a proper choice of the Lagrangian, these theories are equivalence to healthy theories, up to second order in the fields. It is also not clear to what extent such a problem arises in theories with no linear weak-field limit
(for example, in MOND, the Lagrangian is nonanalytic in the fields, in the weak-field limit). Another fact from which we may draw hope, in this connection, is that the modes that appear ghostlike in the linear limit seem to be sourced by the phantom densities. These, in themselves, are not independent of the actual charges; so it may be that a system never actually radiates negative energy to infinity, which is the basic problem with ghosts.
\par
Another subject for further study is the generalization of the concepts here to other NL theories.
The construction of PL theories as described above hinges on the Lagrangian depending nonlinearly only on the first derivatives of the basic DoF. Can a sensible generalization be made to
Lagrangians that depend also (nonlinearly) on the DoF themselves, or on higher derivatives.
For example, for a single scalar, we may consider, instead of genuinely NL theories governed by a field Lagrangian of the form
 \beq \L=-\U(\f_{,\m};\f_{,\m,\n}), \eeqno{bopot}
 PL theories with
\beq \L=-\f_{,\m}\psi_,^{\m}-\f_{,\m,\n}{{{\psi_{,}}^{\m}}_{,}}^{\n}
+\S(\psi_{,\m};\psi_{,\m,\n}). \eeqno{bohop}
This leads to linear, higher-derivative field equations
\beq \Box\psi-\Box^2\psi=q,~~~~~\Box\f-\Box^2\f=\left(\frac{\partial \S}{\partial \psi_{,\m}}\right)_{,\m}-\left(\frac{\partial \S}{\partial \psi_{,\m,\n}}\right)_{,\m,\n}.   \eeqno{lagooo}
However the affinity between the two theories is less clear. Even if we choose $\S(e^\m,\bar e^{\m\n})$ to be the Legendre transform of $\U$, the two theories will not coincide, in general, for 1-D configurations: it is true that in this cases $e_\m=\psi_{,\m}$ for some $\psi$, and $\bar e_{\m\n}=\bar\psi_{,\m,\n}$ for some $\bar\psi$, but the PL theory requires the further constraint $\psi=\bar\psi$.
\par
Other possible generalizations are to PL analogs of generalizations of BI, e.g., certain versions of Dirac-Born-Infeld in Minkowski apace-time, and in (fixed) curved space-time. These involve a set of scalar fields $\Phi^a$,
and in the Minkowski case we take
 \beq \L=E_0^2[1 -(-\|\emn+\omega\Phi^a_{,\m}\Phi^a_{,\n}+\fmn/E_0\|)^{1/2}], \eeqno{murgol}
with summation over double indices.
The PL theory involves also fields $\Psi^a$, and the lagrangian is \beq -\Grad\Phi^a\cdot\Grad\Psi^a-\frac{1}{2}\fmn\Qmn+\S(\Psi^a_{,\m},\qmn) .\eeqno{kureb}

There are also many interesting questions regarding properties of these theories and the the extent of their similarity to their genuinely NL kindred.

\section{\label{appa}Appendix: The short-distance force in Born-Infeld electrostatics}
Consider two charges of the same sign, $q_2$ at the origin, and $q_1$ ($|q_1|\le|q_2|$) at $x=\ell$ on the x axis, in three (space) dimensions. We seek to calculate the force between them in BI electrostatics, in the limit $\ell \rar 0$. The field $\gf$ has three finite critical points: two on the charges, and one in between them where $\gf=0$ at point o. Take as integration surface, $\Sigma$, for calculating the force on $q_1$ from eq.(\ref{forcei}), the ``watershed'' surface, which passes though o, and which separates the field lines ending on the two charges, completed at infinity on the side of $q_1$. On $\Sigma$, $\gf$ is tangent to $\Sigma$. It was shown in \cite{milgrom02} that for a charge distribution of a uniform sign, $\gf$ everywhere points to (or away from) the convex closure of the charge distribution, in our case the segment connecting the two charges. Thus all field lines become radial at large distances $r\gg\ell$, and $\Sigma$ approaches a cone of half opening angle $\theta$.
Apply Gauss theorem to eq.(\ref{muipo}) in the volume within $\Sigma$, which contains only $q_1$. Only the integral at infinity contributes,
and we get that the solid angle subtended by $\Sigma$ at infinity, $\Omega$, is given by $\Omega/4\pi=(1-cos\theta)/2=q_1/(q_1+q_2)$.
\par
Now look at expression (\ref{forcei}) for the force. The contribution to it from infinity is seen to vanish. On the rest of $\Sigma$, $\gf$ is perpendicular to $\dvs$; so we have
\beq \vF=\int_\Sigma u \dvs.\eeqno{forceic}
Since $u\le 0$ the force is repulsive. (Since the tangent to $\Sigma$ always points to the inter-charge segment, the $x$ component of $\dvs$
is everywhere nonpositive.) Divide the integral to the contributions from radii $r\le \kappa\ell$, and $r> \kappa\ell$, with $\kappa\gg 1$ fixed. Since $|u|$ is bounded by $E_0^2$, the first contribution is bounded by $E_0^2(\kappa\ell)^2$, which vanishes in the limit $\ell\rar 0$. Beyond $\kappa\ell$ the two charges are seen approximately as one charge $q_1+q_2$, the field is radial, and to a very good approximation  $u(|\gf|^2)$ may be replaced by its known expression for the spherical, point-charge case:
 \beq u=-E_0^2[1-(1+\frac{y}{E_0^2})^{-1/2}],  \eeqno{cupit}
where $y\equiv [(q_1+q_2)/4\pi r^2]^2$ is the Coulomb field squared, and $r$ is the distance from the origin.
This approximation is arbitrarily good for large enough $\kappa$.
The integral beyond $\kappa\ell$ can now be extended back to the origin with expression (\ref{cupit}) for $u$, again because the integral from the origin to $\kappa\ell$ vanishes in the limit $\ell\rar 0$. The expression for the force in the limit is then the integral (\ref{forceic}), with $u$ from eq.(\ref{cupit}), calculated on the circular-cross-section cone of half opening angle $\theta$, around the $x$ axis.
From symmetry, only the $x$ component is finite, and in the limit $\ell\rar 0$ is
 \beq F(q_1,q_2,\ell\rar 0)\rar 2\pi sin^2\theta E_0^2\int_0^\infty rdr[1-(1+\frac{y}{E_0^2})^{-1/2}]. \eeqno{gumcha}
Integrating, and substituting the expression for $\theta$ in terms of the charges, we get
 \beq F(q_1,q_2,\ell\rar 0)\rar \frac{E_0q_1 q_2}{|q_1+q_2|}.  \eeqno{bueqa}
\par
Following the same calculation in two dimensions gives
\beq F(q_1,q_2,\ell\rar 0)\rar \frac{E_0|q_1+q_2|}{\pi}
sin\left(\pi\frac{q_1}{q_1+q_2}\right). \eeqno{hutatap}
This is the same as the result found in \cite{pryce35}, where two-dimensional BI electrostatics had been considered.
\par
In $D$ dimensions, work in spherical coordinates around the charge axis. The volume element is then
\beq dV=r\^{D-1}sin\^{D-2}\theta sin\^{D-3}\varphi_1...sin\varphi\_{D-3}dr d\theta d\varphi_1...d\varphi\_{D-2} \eeqno{byrewas}
($0\le\theta<\pi$, $0\le\varphi_k<2\pi$). The $D-1$ area on the cone of constant $\theta$ between $r$ and $r+dr$ is
\beq dA=sin\^{D-2}\theta \I\_D r\^{D-2}dr, \eeqno{munshata}
where $\I\_D$ is the integral over the $\varphi_k$ in the expression for $dV$:
$\I\_D\equiv\int sin\^{D-3}\varphi_1...sin\varphi\_{D-3} d\varphi_1...d\varphi\_{D-2}$. This integral also appears in the expression for the $D$-dimensional solid angle, $\a\_D=\I\_D\int_0^\pi sin\^{D-2}\theta d\theta$. Now, the Coulomb field of a point mass $Q$ is
$|\gps|=|Q|/\a\_D r\^{D-1}$. Inserting in the expression for $u$ and integrating, as before, the ratio $\I\_D/\a\_D$ appears, and we find for the short-distance-limit force
 \beq F\rar\frac{E_0|Q|}{D-1}\frac{sin\^{D-1}\theta}{\I^*\_D}, \eeqno{fures}
where $\I^*\_D\equiv \int_0^\pi sin\^{D-2}\theta d\theta=\pi\Gamma(D-1)/2\^{D-2}[\Gamma(D/2)]^2$.
Finally, we express $\theta$, and hence the force, in terms of the charges. Again,
from Gauss's law, the field lines ending at any of the charges, subtend at infinity a solid angle in proportion  to that charge. So,
the opening angle $\theta$ is given by
 \beq \int_0^\theta sin\^{D-2}\beta d\beta=\I^*\_D\frac{q_1}{Q}.  \eeqno{sumna}
To lowest order in $q_1/Q$, eq.(\ref{sumna}) tells us that
$\theta\^{D-1}\approx (D-1)\I^*\_D q_1/Q$, and $F\approx E_0|q_1|$, as expected.
For equal charges $q_1=q_2=q$, $\theta=\pi/2$ and
 \beq F\rar\frac{2\^{D-1}[\Gamma(D/2)]^2}{\pi(D-1)\Gamma(D-1)}E_0|q|. \eeqno{hutras}

I have not been able to fully derive similar results for the PL theory. In this case, there is, generally, no common ``watershed'' surface for both potentials. If we use as integration surface the `watershed' of $\psi$, we get for the force
\beq \vF=\int_\Sigma -(s+\gf\cdot\gps) \dvs+\gps(\gf\cdot \dvs).\eeqno{forceiii}
As before, divide the integral to the contributions for radii below and above $\kappa\ell$. For $\kappa\gg 1$ the fields in the large-radius region become the fields for a point charge $Q$, and $\Sigma$ becomes a radial cone. Thus $\gf$ becomes perpendicular to $\dvs$, so the third term in the integrand contributes negligibly. Also, it is easy to ascertain that for spherical configurations $-(s+\gf\cdot\gps)=u$
(this expression then becomes minus the Legendre transform of $s$, which equals $u$). Thus, the contributions to the force integrals from the region $r>\kappa\ell$ are the same in the two theories for $\kappa\gg 1$. It is also clear that if we use in the integral the fields for the point-mass $Q$, the contribution from the region $r\le\kappa\ell$ vanishes in the limit $\ell\rar 0$, for fixed $\kappa$. However, I have not been able to show that this is also the case for the small-radii contribution to the actual integral. In the BI case, the integrand $u$ is bounded; so this contribution vanishes at least as fast as $\kappa\ell)^2$. If this can be shown to be the case for the PL theory, we would get the same expression for the force. But in the PL theory, the contribution from $r\le\kappa\ell$ may be finite, in which case the forces differ in the two theories.
This remains an open question.

\par
Note, finally, that the above derivation for the BI case can be used to calculate the short-distance force for other configurations involving point charges.
For example, consider $N$ charges, $q_i$, of equal sign, on a segment of the $x$ axis ($i$ increasing in the positive $x$ direction). The force on any $q_k$, in the limit of shrinking configuration, can be calculated by subtracting the integrals over the two `watershed'' surfaces flanking this charge, giving
 \beq F_k\rar \frac{E_0q_k}{Q}(\sum_1^{k-1}q_i-\sum_{k+1}^Nq_i)  \eeqno{makoca}
 ($Q=\sum_1^N q_i$).
(We cannot apply the same to a general point-charges configuration, because we do not know, in general, the asymptotic shape of the `watershed' for each charge.)
\par
Another example is the short-distance force on each charge in a system of $N$ equal charges placed symmetrically on the vertices of a polygon, or a symmetric polytope (cube, tetrahedron). It is based on the fact that the `watersheds' are symmetry surfaces, and so are known.
For example, for $N$ equal charges at the vertices of a regular polygon, $\Sigma$ is made of two half-planes at an angle $2\pi/N$ to each other. So the integration would yield the same value as in he two-charge case, but now the planes make an angle $\pi/N$ with the charge axis. So the result has to be multiplied by $sin(\pi/N)$. Also, in the expression for the asymptotic field we have to take $Nq$ as the total charge.
This gives for the force:
 \beq F=\frac{1}{4}E_0N|q|sin\left(\frac{\pi}{N}\right).  \eeqno{milpar}

For $8$ charges $q$ on the vertices of a cube, we have $\Sigma$ made of three quarter-planes. We then get $F=(3^{1/2}/2)|q|E_0$.
\par
This research was supported, in part, by a center of excellence
grant from the Israel Science Foundation.

\end{document}